\documentclass[aps, prl, twocolumn, superscriptaddress, amsmath,  tightenlines, longbibliography]{revtex4-2}
\usepackage{graphicx}
\usepackage{dcolumn}
\usepackage{bm}
\usepackage{bbm}
\usepackage{framed} 
\usepackage{tcolorbox}
\usepackage{braket}
\usepackage{float}	
\usepackage{chapterbib}
\usepackage{amsthm,amsmath,amssymb}
\usepackage{mathrsfs}
\usepackage{amsfonts}	
\usepackage{epsfig}
\usepackage{multirow}	
\usepackage{booktabs}
\usepackage{color}
\usepackage{xcolor}

\newcommand{\nvec}[1]{\textbf{#1}}

\newcommand{\im}{i}
\newcommand{\ii}{\im}

\newcommand{\e}{e}

\def\ie{{\it \textit{i.e.,~}\/}}

\begin{document}
\preprint{APS/123-QED}

\title{Observing the nodal-line conversion determined by the relative homotopy}
\author{Maopeng Wu}
	\affiliation{State Key Laboratory of Tribology, Department of Mechanical Engineering,\\
Tsinghua University, Beijing 100084, China}	
\author{Mingze Weng}
	\affiliation{State Key Laboratory of Tribology, Department of Mechanical Engineering,\\
Tsinghua University, Beijing 100084, China}	
\author{Qian Zhao}%
 	\email{zhaoqian@tsinghua.edu.cn}
 	\affiliation{State Key Laboratory of Tribology, Department of Mechanical Engineering,\\
Tsinghua University, Beijing 100084, China}
\author{Yonggang Meng}%
 	\affiliation{State Key Laboratory of Tribology, Department of Mechanical Engineering,\\
Tsinghua University, Beijing 100084, China}
\author{Ji Zhou}
	\affiliation{State Key Laboratory of New Ceramics and Fine Processing,\\
School of Materials Science and Engineering, Tsinghua University, Beijing 100084, China}	
\date{\today}

\begin{abstract}
Directly identifying the non-Abelian nodal-line semimetals (NASM) is quite challenging because nodal-line semimetals typically do not possess topologically protected boundary modes. Here, by reconstructing the correspondence between the bulk states of Hermitian systems and circuit voltage modes through gauge scale potential, the temporal topolectrical circuits (TTC) for evidencing NASM are proposed. Following the logical progress of discovering NASM, we start by demonstrating the relative homotopy group of  two-band models using TTC, which can faithfully determine the conversion rules between the nodes in and out of the non-local-symmetry invariant subspace. Next, we 
show that those rules dramatically change with the consideration of the additional band, historically leading to the arising of the NASM. Also, we demonstrate the unique non-Abelian constrained nodal configuration - earing nodal lines. Our results established NASM for further investigating topological line degeneracies, and proposed TTC will be a versatile platform for exploring nodal-line semimetals.
\end{abstract}

\maketitle
\textit{Introduction---}With symmetries protected (\ie PT symmetry, chiral, and mirror symmetries), two energy bands of the nodal line semimetals can intersect each other along the closed curve in the three dimensional Brillouin zone \cite{aau8740,PhysRevB.84.235126,Fang_2016}. Those degeneracy curves are called the nodal lines (NL) \cite{PhysRevB.84.235126} and they behave as various shapes, such as rings \cite{RevModPhys.90.015001}, chains \cite{bzduvsek2016nodal}, knots \cite{PhysRevB.96.201305}, and links \cite{PhysRevB.96.041103}. Due to symmetry breaking, nodal line semimetals can converse to Weyl \cite{PhysRevLett.119.036401} or Dirac semimetals \cite{PhysRevLett.115.036807}. Recently, homotopy theory has been applied to understand the band intersection \cite{PhysRevB.96.155105, PhysRevLett.121.106402}, including complete nodal classification and node transfer. Beyond the established “tenfold way” methods, one of the unique nodal classification is the non-Abelian nodal-line semimetals (NASM) \cite{aau8740,PhysRevB.101.195130}, which exhibit braiding topological structures \cite{bouhon2020non} and trajectory-dependent node transfer \cite{jiang2021experimental,guo2021experimental}. Notably, like normal NL semimetals, NASM does not have protected boundary modes that are usually attended by topological phases \cite{Fang_2016,RevModPhys.90.015001}, preventing NASM from being deterministic.\par

The physical realization of non-Abelian NL (such as linked “earrings” NL) and, mostly, experimentally characterizing them in the 3D Brillouin zone is exceptionally  challenging. The tight-binding model is widely adopted to describe the band structure in the theory of semimetals \cite{RevModPhys.90.015001}, where broken time-reversal symmetry and spin-orbits coupling are usually involved. But localized orbits are not desired bases for optics, and adding the model-required ingredients, such as spin-orbit coupling, is extremely difficult (even impossible). Experimentally, angle-resolved transmission spectroscopy (\ie photoemission transmission for the electronic band, parameter transmission optical band \cite{yan2018experimental}) and Fourier-transformed field scan (widely used in optics \cite{PhysRevLett.125.033901} and acoustic \cite{jiang2021experimental} metamaterial) are the primary approaches to character the band structure, as well as the band degeneracies, of the (or artificial) material. However, neither of them is appliable to verify the NL because the former is very limited in momentum resolution \cite{Fang_2016}, and the latter requires field distribution that is not accessible in three-dimensions. The site-resolved topological circuits or topolectrical circuits \cite{PhysRevX.5.021031,lee2020imaging} may provide a manageable platform to identify the NASM since the circuit network has a one-to-one correspondence to a tight-binding model and the required ingredients are possible to realize. However, the general theory to demonstrate NASM using circuits is lacking.\par

Here, we propose the temporal topolectrical circuits (TTC) to demonstrate the non-Abelian nodal line semimetals with Hermiticity. Contrary to the existing topological circuits, we show that the gauge scale potential in the circuit lattice is crucial in the correspondence between the eigenmodes of the Hamiltonian and the voltage modes of TTC. And based on that, further selective excitation of the voltage modes enables us to acquire the band information along with the specific crystal orientation or on the specific crystal face, thus making TTC momentum-resolved. \textit{The above features of TTC allow us to identify the NASM directly.} Our 
 demonstration follows the mathematical description of NASM with homotropy theory: (i) First, in the two-band model, we show that mirror-related Weyl points with opposite chirality convert into NL at the mirror-invariant plane rather than annihilating. That conversion is ruled by the relative homotopy group $\pi_2(M_2,X_R)$. (ii) Next, in the three-band model, we demonstrate that crossing points of intersecting NL transfer to those formed by another pair of bands. The above transfer is out of the capacity of the two-band relative homotopy group $\pi_1(M_{1,2},X_R)$, hence historically leading to the non-Abelian description of band topology. (iii) we identify the unique earring NL constrained by the non-Abelian property of $\pi_1(M_3)$, which has not been previously reported. \par
\begin {figure}[h]
\centering
\includegraphics[width=9cm]{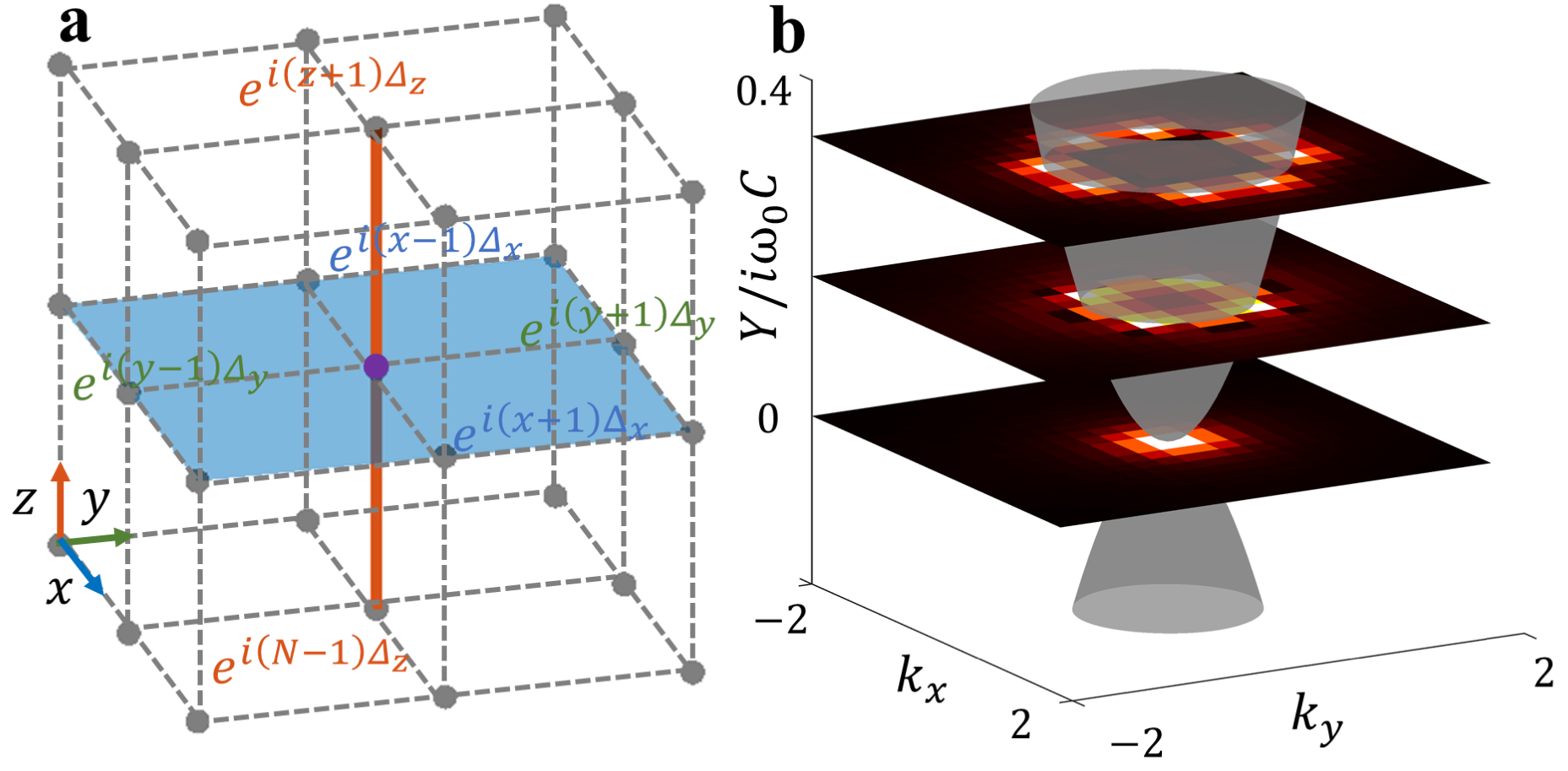}
\caption{Momentum-resolved TTC due to selective excitation and enabled FTFS owing to $YI$. (a) The purple dot represents the point source that excites the bulk states with identity eigenenergy. Hence the constructed band after Fourier transform is the iso-energy contour. The orange line is the line source, along which neighbor sites have a phase difference $\Delta _z$. That carefully designed line source will excite the bulk states with identical eigenenergy and wavevector $k_z =\Delta _z$. Thus band information on the momentum plane $\mathbf{k} = (k_x,k_y,\Delta _z)$ is accessible. Similarly, band information in $k_z$ direction is also available with face source (bule face). (b) Quadratic degeneracy formed by the Weyl node colliding (with respect to inset of Fig. 2b). The planes with scaled colors show the FTFS results.}
\label{fig1}
\end {figure}
\textit{Momentum-resolved TTC---} We show the admittance and voltage modes correspond to the Hamiltonian and its field coherences, respectively. When the lattice system described by $H(\mathbf{k})$ is isolated to the environment, its density matrix $\rho$ is governed by Von Neumann equation $\mathrm{d}\rho /\mathrm{d}t = -i\left [ H, \rho \right ]$. If we employ the field coherences $\phi_{\mathbf{k},i}\left ( t\right )=\left \langle a_{\mathbf{k}, i}\left ( t\right )\right \rangle = Tr\left [ a_{\mathbf{k},i}\rho\left ( t\right )\right ]$ to monitor the time evolution of $\rho$, then
\begin{equation}
\phi_{\mathbf{k},n}\left ( t\right )= \Phi_{\mathbf{k},n}e^{iE_n(\mathbf{k})t-i\mathbf{k}\cdot \mathbf{r}},
\label{psi}
\end{equation}
where $\Phi_{\mathbf{k},n}$ and $E_n(\mathbf{k})$ are the eigenvectors and eigenvalues of $H$, respectively. 
On the other hand, with Bloch theory, the admittance matrix of the circuit in a lattice structure can be written as $J\left ( w, \mathbf{k}\right )  = YI + J_{0}\left ( w, \mathbf{k}\right ) $, where the diagonal elements $YI$ (so-called self admittances) are separated and preset to be indentical. Through time-domain
analysis, the node voltage $V_{\nvec{k},0}(t)$ that changes over time are \cite{wu}
\begin{align}
    V_{\nvec{k}}(t) = \nvec{V}_{\nvec{k},0} \, \e^{\ii \omega(\nvec{k}) t- \ii \nvec{k} \cdot \nvec{r}}, 
\label{volOverTime}
\end{align} 
where $\nvec{V}_{\nvec{k},0}$ are the eigenvectors of $J\left ( w, \mathbf{k}\right )$ with respect to the \textit{zero-value eigenvalues}, $w$ is the resonant frequency of the circuit. \par 
The above equation reminds us of Eq. 1, and indicates that the response of the node voltage can emulate the evolution of the field coherences, as long as $J\left ( w, \mathbf{k}\right )$ is appropriately configured to be of the same form as that of $H(\mathbf{k})$. 
To do so, the circuit has to operate at a \textit{fixed} frequency 
 since $J\left ( w, \mathbf{k}\right )$ is frequency-dependent.
But the frequency dimension in normal meta-materials usually corresponds to eigen-energy of a solid. So to make the circuit temporal, one needs to find out a circuit parameter that corresponds to the eigen-energy.
Besides, the voltage response $V_{\nvec{k}}(t)$ is modeled by $\nvec{V}_{\nvec{k},0}$ with vanishing eigenvalue, while measuring $\nvec{V}_{\nvec{k},j\ne 0}$ with a non-vanishing eigenvalue is necessary to characterize the complete band structure. We show the above two limitations can be solved by self-admittances $YI$.\par

The $YI$ plays a similar role to the onsite scalar potential in the tight-binding model. Tuning the onsite potentials globally will shift the band by a constant in the energy dimension, but it doesn't cause any physical effect
 since it can be gauged out. But in TTC, it can control the eigenvectors  $\nvec{V}_{\nvec{k},0}$ that correspond to zero-value eigenvalues without changing the operating frequency (more details see \cite{wu}). In other words, one can select the evolution of the bulk node voltage modes with $YI$. Also, parameter scanning of the bulk modes with $Y$ enables us to character band of the circuit (via a Fourier transform, Fig.1b). 

The TTC are both site and momentum resolved due to their circuit nature. Different from distributed parameter metamaterials, bulk voltage modes inside the 3D TTC are easily accessible, thus leaving TTC site-resolved. 
Also, the eigenmodes of TTC can be selectively excited through careful-designed space-distribute excitation, making it momentum resolved. Indeed, according to Breit-Wigner decomposition, the interaction $W_{\mathbf{k},exc}$ between the circuit modes $V_{\mathbf{k}}$ and outside current excitation  $I_c e^{\ii \omega t}$  reads
\begin{equation}
W_{\mathbf{k},exc}=\sum_{c}  I_ce^{i\mathbf{\Delta }(\mathbf{r_c})\cdot \mathbf{r_c}}V_{\mathbf{k} }^{*}(\mathbf{r_c} )
\end{equation}
where $\mathbf{\Delta }(\mathbf{r_c})$ are phases of excitation, which are dependent on their position $\mathbf{r_c}$ in lattice. $W_{\mathbf{k},exc}=0$ suggests that corresponding modes can not be excited. Figure 1a shows the source configuration that can excite the eigenmodes along the specific crystal orientation or on the specific crystal face (more details see \cite{wu}). \par
\textit{The general relative homotopy description of band nodes can be found in \cite{aau8740, PhysRevLett.121.106402}. Here, we use the above theory to analyze our circuit models and validate them with 
simulation results.}\par
\begin {figure}[h]
\centering
\includegraphics[width=\linewidth]{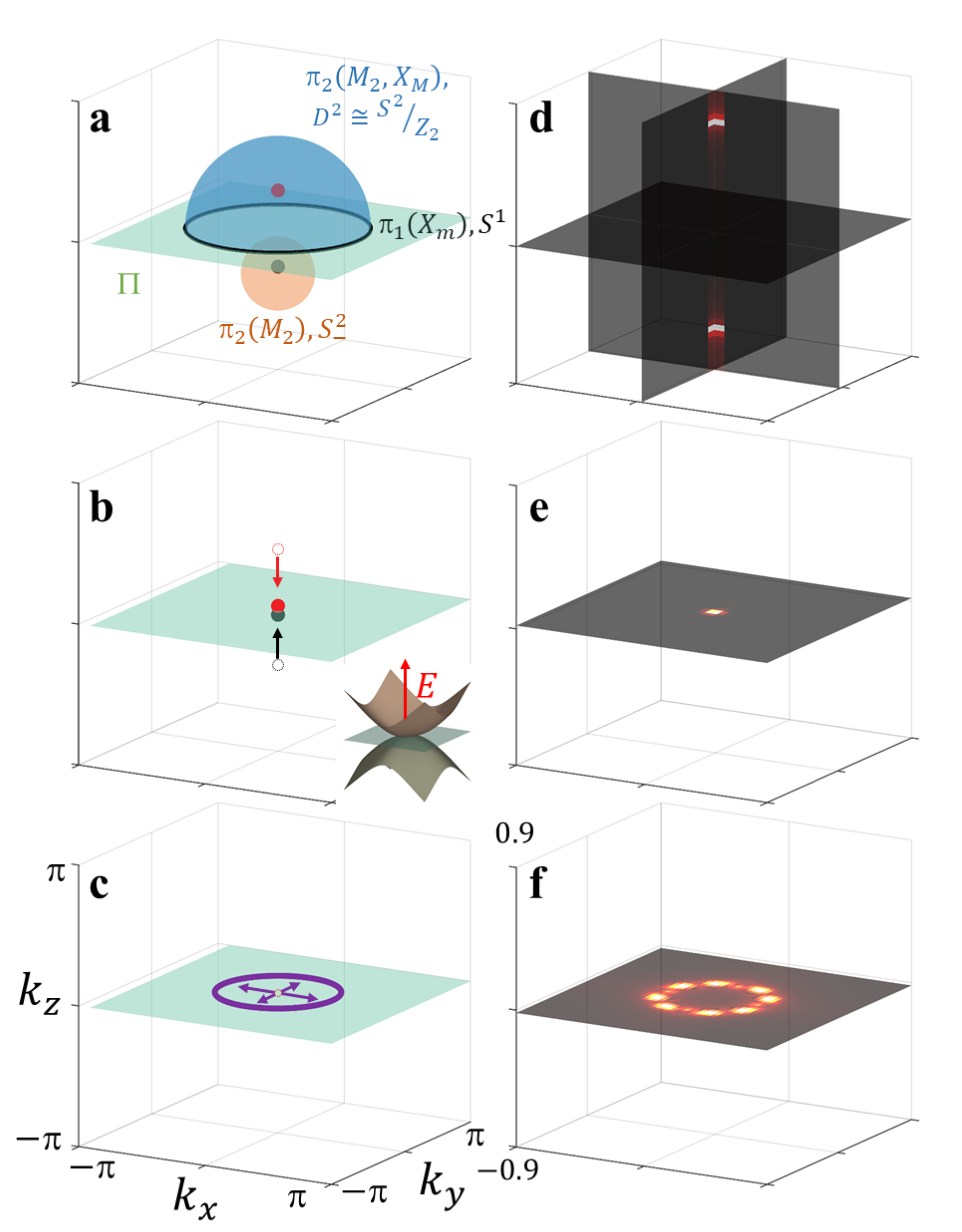}
\caption{Weyl node conversion in circuits determined by the relative homotopy group $\pi_2(M_2, X_M)$. The green plane denotes the mirror-invariant plane $\Pi$. We don't plot the NL on $k_z = \pi$ plane for clearness since it can be evaded by selective excitation. (a) Mirror-related Weyl nodes with opposite chirality. We assign the black node surrounded by the orange sphere $S^2$ to $n = -1$. The blue hemisphere is the $D^2$ used in $\pi_2(M_2, X_M)$ and $D^2 \cap \Pi = S^1$ (black ring), which is constant in $M_2$.  $A = 0.9$. (b) Weyl node colliding on $\Pi$ forming a quadratic degeneracy (inset). $A = 0.6$. (c) Weyl nodes convert to a nodal line (purple) rather than annihilating due to non-trivial $n_r=1$.  $A = 0.2$. (d)-(f) Numerical results (via FTFS) from \textit{LT spice} with respect to (a)-(c)}
\label{fig2}
\end {figure}
\textit{Two-band model---}Our demonstration starts with the two-band models. The purpose is twofold: (i) we would like to demonstrate that the relative homotopy group captures the two-band conversion rules between the in-plane and out-of-plane (the absence of the 3-band historically leads to the non-Abelian classification), (ii) and we want to show the excellent facilities of the momentum-resolved TTC in 3D, which all set a theoretical and experimental stage for later non-Abelian sections. For concreteness, the circuit model considered here is

\begin{equation}	
\begin{aligned}
(jw_0C)^{-1}J_1(\mathbf{k} ) = & \mathrm{sin} k_z \left (\mathrm{sin}k_x\sigma_x + \mathrm{sin}k_y\sigma_y\right )+ \\
&[-1-0.8(\mathrm{cos}k_x+\mathrm{cos}k_y)-A\mathrm{cos}k_z]\sigma _z
\end{aligned}
\end{equation}
where $\sigma_{i=x,y,z}$ are the Pauli matrices and $A$ is the tuning parameter. The above model does not possess any lock-in symmetry (\ie $T$ reverse symmetry). Note that  to break $T$ symmetry 
 we employ the negative impedance converters \cite{wu}. Therefore projected model $\mathbb{R}\times (\mathbb{R}^3 \backslash \left \{ \mathbf{0} \right \} )$ (by $J'(\mathbf{k} ) = I - 2\sum _n\ket{u_{n}(\mathbf{k})}\bra{u_{n}(\mathbf{k})}$, $u_{n}(\mathbf{k})$ are occupied states) belongs to a topological space $M_2$: $J(\mathbf{k})\subset M_2 \cong S^2$, which is classified as class A in \cite{PhysRevB.96.155105}. The embeddings of the $p$-dimensional sphere $S^p$ in $M_2$ formulate the homotopy group $\pi_p(M_2)$. The non-trivial element in the group indicates the embedding cannot be continuously deformed to a point, therefore gapless nodes must exist inside the disc $D^{p+1}$ (boundary of the $S^p$, $\partial D^{p+1} =S^p$), \ie $\pi_{p=2}(M_2)=\mathbb{Z}$ and non-trivial element suggests Weyl points. For our model with $A=0.9$, the orange sphere $S^2_-$ in Fig. 2a cannot shrink to a point because of the Weyl node it encloses. And the winding number $n=-1$ on $S^2_-$ suggests that the chirality of this Weyl node is $-1$.\par
The circuit model is mirror-symmetric $\sigma_z J(k_x,k_y,k_z)\sigma_z = J(k_x,k_y,-k_z)$, where the $\sigma_z$ is the representation of the symmetry. Inside the two mirror-invariant planes $\Pi$ of the Brillouin zone $\Pi = (0,0,0/\pi)$, we have the subspace $X_m$ of $M_2$, whose elements are commutate with $\sigma_z$ for $\forall \  \mathbf{k}\in  \Pi$. The continuous maps $J: D^p \rightarrow M_2$ with $J(S^1)\subset X_m$ lead to the topological classification: relative homotopy group $\pi_p(M_2,X_m),\ p=1,2,...$, where the embedding $S^p$ in $\pi_p(M_2)$ is a disk $D^p$ now since $D^p$ is identical to the upper hemisphere of the mirror-symmetric $S^p$ that fully contains the topological information.\par

Our circuit model exhibits two $m_z$-related Weyl nodes with opposite chirality when $A>0.6$ (Fig. 2a and 2d). The winding number $n$ on the sphere that encloses both points is trivial $n=0$, meaning that two nodes can either annihilate (bandgap open after that) or convert to other types of nodes (such as nodal lines) with adiabatically tuning. Annihilating or not is controled by  $\pi_2(M_2,X_m)= \mathbb{Z}$. The image of $S^1$ in $\Pi$ (black circle in Fig. 2a) is a constant in $M_2$, so $n_r$ of $\pi_2(M_2,X_m)$ on $D^2$ (blue semisphere in Fig. 2a) is equal to $n$ on the sphere that sololy includes the positive Weyl point $n_r = n =1$, whereas the nontrivial $n_r$ corresponds to the latter conversion case after the node colliding. Figure 2b and 2e show the $m_z$-related Wely points collide when $A=0.6$, forming a double Wely node (or Euler class \cite{jiang2021experimental}) with quadratic degeneracy (inset of Fig. 2b and Fig. 1b). Decreasing the parameter continually,  the double Wely node will stretch into an $m_z$-protected nodal line (Fig. 2c and 2f). 
\par
\begin {figure}[htbp]
\centering
\includegraphics[width=\linewidth]{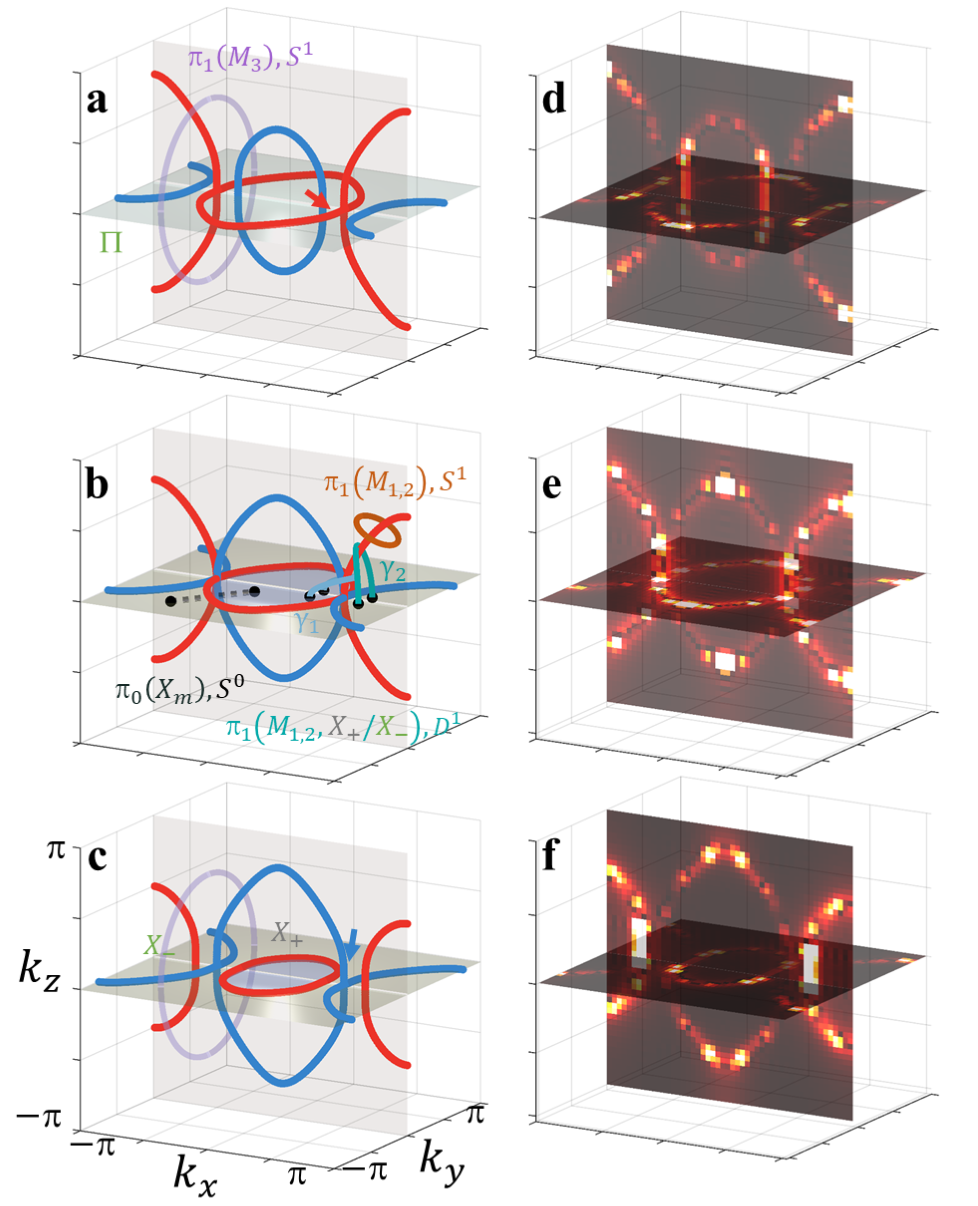}
\caption{Relative homotopy group $\pi_1(M_2, X_{+/-})$ protecting CP is absent and CP transfer for 3-band model. The red and blue lines denote the NL formed by lower (upper) two bands resp.  (a) CP intersected by the occupied and unoccupied bands. The purple loop is the $S^1$ used in $\pi_1(M_3)$ and winding on it is $n_{\Gamma}=i^2=-1$. $B=0.5$. (b) CP survive when $B=0$. The orange loop is the $S^1$ used in $\pi_1(M_{1,2})$ and $\gamma_{1,2}$ is the $D^1$ used in $\pi_1(M_{1,2}, X_{+/-})$. (c) NL distangle due to $\pi_1(M_{1,2}, X_{+})=0$. $X_{+/-}$ are separated by the red NL, in which $\partial \gamma_{1,2}$ lie. The winding on $S^1$ of $\pi_1(M_{3})$ is $n_{\Gamma}=k^2=-1$. (d)-(f) Numerical results with respect to (a)-(c).}
\label{fig3}
\end {figure}
\textit{Three-band model---}In this section, we demonstrate that $\pi_1(M,X_m)$ description of the two-band is absent for the three-band model. Compared to $\pi_2(M,X_m)$ that captures the nodal point conversion, $\pi_1(M,X_m)$ determines the NL convention rules. The in-plane NL and the convented out-of-plane NL intersect at the cross points (CP), so one can alternatively understand $\pi_1(M,X_m)$ protects the CP. To demonstrate the topological stability of CP modified by the additional band, the concrete circuit lattice model considered here is 
\begin{equation}
\begin{aligned}
&(jw_0C)^{-1}J_3(\mathbf{k} ) = 
\left ( \mathrm{cos}k_x -1 \right ) \left \{ S_x,S_y \right \} + \\& \left ( \mathrm{cos}k_z -1 \right ) \left \{ S_z,S_x \right \}
-B\mathcal{S}_1 +\mathcal{S}_2\mathrm{cos}k_y -\mathcal{S}_3\mathrm{cos}k_y ,
\end{aligned}
\end{equation}
where $3\mathcal{S}_i = w^{i-1}\Gamma+h.c.-I  $, $\Gamma = S_x^2+wS_y^2+w^2S_z^2 $, $w= e^{2\pi i /3}$. $S_i$ is the angular momentum matrices for spin 1.  The above model preserves the inversion symmetry $P$, time-reversion symmetry $T$, and the combined symmetry $PT$. Up to a gauge, the $PT$-symmetry enforces $J_2(\mathbf{k} ) $ to be real (classified as nodal class AI \cite{PhysRevB.96.155105}). We remark that the continuum $T$-breaking models in \cite{aau8740} can be easily generalized to lattice models. But $T$-preserving models are more practical and feasible for experimental realization. A similar projection procedure applied to  $J_2(\mathbf{k} ) $ leads to the $J_2(\mathbf{k} ) \subset M_{1,2}\cong \mathbb{R}P^2$, where we assume that the first band is the occupied band and the others are unoccupied bands. $\pi_1(M_{1,2})=\mathbb{Z}_2$ implies that NL threads $S^1$ (a loop) if the winding number on $S^1$ is non-trivial, \ie the orange loop encloses the red NL in Fig. 3b.  \par

The circuit model doesn't respect mirror symmetry, such as $z$-mirror represented by $m_z =\mathrm{diag}(1,1,-1) $,  but the expand near $\mathbf{k_0} =(0,0,\pm \pi/2)$ does,  namely
\begin{equation}
\begin{aligned}
&\lim_{\mathbf{k}  \to \mathbf{k_0} } J_2(\mathbf{k}-\mathbf{k_0})=J_{eff}(\mathbf{\kappa }) \\
&\sigma_z J_{eff}(\kappa_x,\kappa_y,\kappa_z)\sigma_{z} = J_{eff}(\kappa_x,\kappa_y,-\kappa_z).
\end{aligned}
\end{equation}
Since the behavior of NL and CP are fully determined by $J_{eff}$ (just like topology property is determined by the effective Hamiltonian close to Dirac points), hereafter, we focus on $J_{eff}$. The $m_z$-symmetric subspace $X_m\subset M_{1,2}$ consists of two disjoint components $X_m=X_+\mathbf{\sqcup} X_-$ separated by in-plane red NL (shade with purple and green in Fig. 3c), which are homotopic to $X_-\simeq \mathrm{point} $ and $X_+\simeq S^1 $.\par

The $\pi_1(M_{1,2}, X_{+/-})$  is incapable of stabilizing CP. This follows that CP can be disentangled by moving out-of-plane NL to $X_+$ without altering the topological structure. For $B=0.5$ and $0$, $J_{eff}(\mathbf{k} )$  exhibits the intersecting NL formed by the occupied and unoccupied bands (illustrated by the red line in Fig. 3a and Fig. 3d). Here, the upper NL denoted by the blue can be temporarily neglected since the upper two bands are flatted as a single trivial band. The full description, including the upper band, requires multi-band homotopy (\cite{aau8740}, also see next section). The $D^p$ required in the definition of $\pi_p(M_{1,2}, X_{+/-})$ for $p=1$ is an open path, \ie $\gamma_{1,2}$ in Fig. 3b. To endow $\gamma_{1,2}$ non-trivial group structure, the endpoints $\partial \gamma_{1,2}$ of $\gamma_{1,2}$ have to lie inside the same region, \ie $\partial \gamma_{1,2}$ locate inside $X_{+/-}$ respectively. Accordingly, $\pi_1(M_{1,2}, X_{+/-})=0/\mathbb{Z}_2$. The trivial group structure of $\pi_1(M_{1,2}, X_{+})$ implies that there is an $m_z$-symmetric deformation of $J_{eff}(\mathbf{k} )$ that permits $\gamma_1$ contract to a point. Hence, the CP in Fig. 3a and Fig. 3b can be removed by pushing out-of-plane NL to $X_{+}$  ($B=-0.5$, shown in Fig. 3c and 3f). Applying a similar analysis on $\gamma_2$, one can deduce that pushing NL to $X_{-}$  is forbidden.\par

The additional presence of particle-hole symmetry would pin the NL at the same eigenvalue. That property doesn't play an essential role in forming NL but will significantly facilitate the experiments. This roots in that FTFS measures the band information with same eigenvalue, and one can obtain the NL structure through a single measurement, \ie model of Eq. 4. If the model is not particle-hole symmetric, one has to parameter scan the iso-eigenvalue contour, and scrambles the measured information up to construct NL. The data from the above brute force method always accompanies a lot of noise due to bulk states. Here, we avoid this by partially flatting the central second band (through adding a bulk dispersion) \cite{wu}.\par

\begin {figure}[htbp]
\centering
\includegraphics[width=\linewidth]{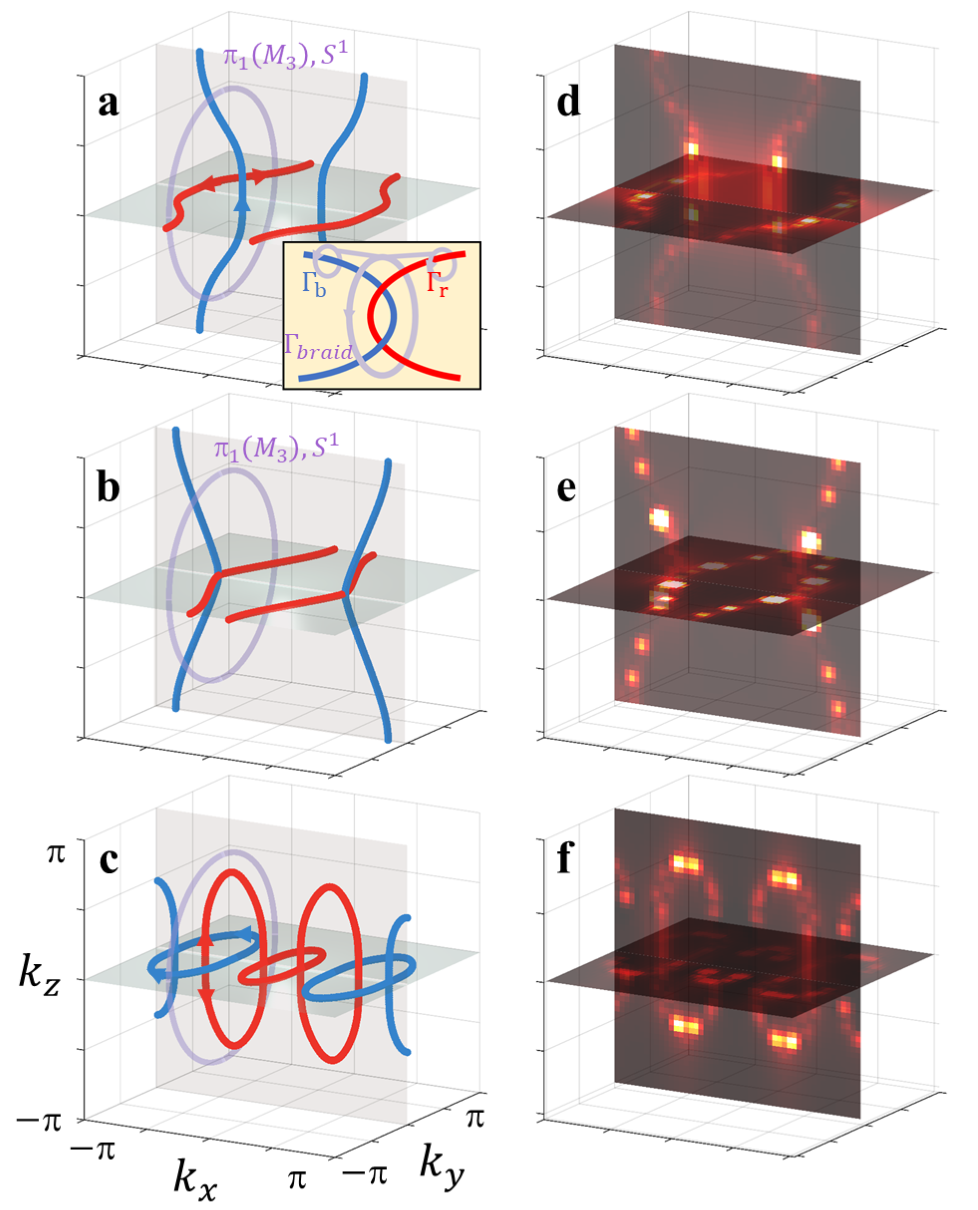}
\caption{Non-Abelian nodal lines and earing nodal structure. The purple loop encircles the braided NL formed by the upper (blue) and lower (red) bands. Arrows denote the orientation of the NL. (a) Tangled NL. $C=1.3$. The inset shows the winding number formula on the purple loop:  $\Gamma_{braid}\sim \Gamma_{R}\Gamma_{B}^{-1}\Gamma_{R}\Gamma_{B}^{-1}$. (b) NL of different types move towards each other, $C=0.5$ (c) NL does not cross each other owing to the quaternion charge $n_{\Gamma}=-1$ on the purple loop,  resulting in the earing NL. $C=-0.3$. (d)-(f) Numerical results with respect to (a)-(c).}
\label{fig4}
\end {figure}
\textit{Earring NL---} However, as we 
observed, the CP doesn't vanish entirely for $B=-1$. Instead, they transfer to another type of intersection (blue NL) formed by the upper bands, as shown in Fig.3c.  To show the non-Abelian nature poses a constrain on NL, the concrete circuit model reads 
\begin{equation}
\begin{aligned}
&(jw_0C)^{-1}J_3(\mathbf{k} ) = 
\left ( \mathrm{cos}k_x -1 \right ) \left \{ S_x,S_y \right \} +  \left ( \mathrm{cos}k_z -1 \right ) \\& \left \{ S_z,S_x \right \}
-2\mathrm{cos}k_y\mathcal{S}_1 +[0.5\mathrm{cos}(2k_y) +C](\mathcal{S}_2-\mathcal{S}_3).
\end{aligned}
\end{equation}
$J_3$ is also PT-symmetric, and the effect model near $\mathbf{k_0}$ preserves $m_z$ and $m_x$ symmetry. The spectral flatting projection that contains the topological information of the third band is $J_{3}'(\mathbf{k} )={\textstyle \sum_{j}^{N=3}} j\ket{u_{\mathbf{k}}^j}\bra{u_{\mathbf{k}}^j}$, where $\ket{u_{\mathbf{k}}^j}$ is the j-th Bloch states. $J_{3}(\mathbf{k} )\subset M_3\equiv \mathrm{SO} (3)/\mathrm{D_2} $ and $\pi_1(M_3)=\pi_0(\mathrm{D_2} )=\left \{ \pm 1, \pm i,\pm j,\pm k\right \} $ (represented by the quaternion group) with $i^2=j^2=k^2=-1$.
 The element $i$ ($k$) in the group corresponds to that $S^1$ encircles NL formed by the lower (upper) two bands, while $k$ encircles both and the sign $\pm$ corresponds to the NL orientation. The non-Abelian property manifest itself as the anticommuting relation $ik=-ki=-j$, \ie $ik=-ki$ implies that orientation reverses if NL passes under NL of the other type (red and blue arrows in Fig. 4a-4c).\par
The element $-1$ captures the entanglement between the NL of the same type or that of different types. For instance, in the former case, CP transfer in the model of $J_2$ (shown in Fig. 3a-3c) corresponds to the winding conservation on the path, which reinterprets winding $-1$ from $-1=i^2$ to $-1=k^2$. The latter case will be 
investigated in the model of $J_3$.  For $C=1.3$, the model has extended and detangled NL, and the winding number on the purple path is $n_{\Gamma } = i\cdot (-k)\cdot (-i)\cdot k=-1$ (Fig. 4a and 4d, inset of Fig. 4a shows the $n_{\Gamma }$ formula of braided NL on the path). For $C=0.5$, NL of different types touches each other (Fig. 4b and 4e). However, further adiabatically tuning $C$ below $0.5$ does not force NL move across each other, since nontrivial winding $-1$ forbids that. In fact, they present as tangled NL with an earing structure ($C=-0.3$, Fig. 4c and 4f). \par 
\textit{Conclusion---} Directly evidencing the topological property in 3D that does not feature surface states is quite challenging. Here, we propose a site-resolved and momentum-resolved circuit, which can serve as a general experimental platform. As an example of that, we 
demonstrate the band degeneracy in semimetal and further non-Abelian semimetal. Our circuit may inspire metamaterial reverse design (such as that in acoustic \cite{PhysRevLett.114.114301} and EM \cite{ZHAO200960} frequency range) since those can be simplified as 
the lumped parameter electrical circuit.


\begin{thebibliography}{22}%
\makeatletter
\providecommand \@ifxundefined [1]{%
 \@ifx{#1\undefined}
}%
\providecommand \@ifnum [1]{%
 \ifnum #1\expandafter \@firstoftwo
 \else \expandafter \@secondoftwo
 \fi
}%
\providecommand \@ifx [1]{%
 \ifx #1\expandafter \@firstoftwo
 \else \expandafter \@secondoftwo
 \fi
}%
\providecommand \natexlab [1]{#1}%
\providecommand \enquote  [1]{``#1''}%
\providecommand \bibnamefont  [1]{#1}%
\providecommand \bibfnamefont [1]{#1}%
\providecommand \citenamefont [1]{#1}%
\providecommand \href@noop [0]{\@secondoftwo}%
\providecommand \href [0]{\begingroup \@sanitize@url \@href}%
\providecommand \@href[1]{\@@startlink{#1}\@@href}%
\providecommand \@@href[1]{\endgroup#1\@@endlink}%
\providecommand \@sanitize@url [0]{\catcode `\\12\catcode `\$12\catcode
  `\&12\catcode `\#12\catcode `\^12\catcode `\_12\catcode `\%12\relax}%
\providecommand \@@startlink[1]{}%
\providecommand \@@endlink[0]{}%
\providecommand \url  [0]{\begingroup\@sanitize@url \@url }%
\providecommand \@url [1]{\endgroup\@href {#1}{\urlprefix }}%
\providecommand \urlprefix  [0]{URL }%
\providecommand \Eprint [0]{\href }%
\providecommand \doibase [0]{https://doi.org/}%
\providecommand \selectlanguage [0]{\@gobble}%
\providecommand \bibinfo  [0]{\@secondoftwo}%
\providecommand \bibfield  [0]{\@secondoftwo}%
\providecommand \translation [1]{[#1]}%
\providecommand \BibitemOpen [0]{}%
\providecommand \bibitemStop [0]{}%
\providecommand \bibitemNoStop [0]{.\EOS\space}%
\providecommand \EOS [0]{\spacefactor3000\relax}%
\providecommand \BibitemShut  [1]{\csname bibitem#1\endcsname}%
\let\auto@bib@innerbib\@empty
\bibitem [{\citenamefont {Wu}\ \emph {et~al.}(2019)\citenamefont {Wu},
  \citenamefont {Soluyanov},\ and\ \citenamefont {Bzdušek}}]{aau8740}%
  \BibitemOpen
  \bibfield  {author} {\bibinfo {author} {\bibfnamefont {Q.}~\bibnamefont
  {Wu}}, \bibinfo {author} {\bibfnamefont {A.~A.}\ \bibnamefont {Soluyanov}},\
  and\ \bibinfo {author} {\bibfnamefont {T.}~\bibnamefont {Bzdušek}},\ }\href
  {https://doi.org/10.1126/science.aau8740} {\bibfield  {journal} {\bibinfo
  {journal} {Science}\ }\textbf {\bibinfo {volume} {365}},\ \bibinfo {pages}
  {1273} (\bibinfo {year} {2019})}\BibitemShut {NoStop}%
\bibitem [{\citenamefont {Burkov}\ \emph {et~al.}(2011)\citenamefont {Burkov},
  \citenamefont {Hook},\ and\ \citenamefont {Balents}}]{PhysRevB.84.235126}%
  \BibitemOpen
  \bibfield  {author} {\bibinfo {author} {\bibfnamefont {A.~A.}\ \bibnamefont
  {Burkov}}, \bibinfo {author} {\bibfnamefont {M.~D.}\ \bibnamefont {Hook}},\
  and\ \bibinfo {author} {\bibfnamefont {L.}~\bibnamefont {Balents}},\ }\href
  {https://doi.org/10.1103/PhysRevB.84.235126} {\bibfield  {journal} {\bibinfo
  {journal} {Phys. Rev. B}\ }\textbf {\bibinfo {volume} {84}},\ \bibinfo
  {pages} {235126} (\bibinfo {year} {2011})}\BibitemShut {NoStop}%
\bibitem [{\citenamefont {Fang}\ \emph {et~al.}(2016)\citenamefont {Fang},
  \citenamefont {Weng}, \citenamefont {Dai},\ and\ \citenamefont
  {Fang}}]{Fang_2016}%
  \BibitemOpen
  \bibfield  {author} {\bibinfo {author} {\bibfnamefont {C.}~\bibnamefont
  {Fang}}, \bibinfo {author} {\bibfnamefont {H.}~\bibnamefont {Weng}}, \bibinfo
  {author} {\bibfnamefont {X.}~\bibnamefont {Dai}},\ and\ \bibinfo {author}
  {\bibfnamefont {Z.}~\bibnamefont {Fang}},\ }\href
  {https://doi.org/10.1088/1674-1056/25/11/117106} {\bibfield  {journal}
  {\bibinfo  {journal} {Chinese Physics B}\ }\textbf {\bibinfo {volume} {25}},\
  \bibinfo {pages} {117106} (\bibinfo {year} {2016})}\BibitemShut {NoStop}%
\bibitem [{\citenamefont {Armitage}\ \emph {et~al.}(2018)\citenamefont
  {Armitage}, \citenamefont {Mele},\ and\ \citenamefont
  {Vishwanath}}]{RevModPhys.90.015001}%
  \BibitemOpen
  \bibfield  {author} {\bibinfo {author} {\bibfnamefont {N.~P.}\ \bibnamefont
  {Armitage}}, \bibinfo {author} {\bibfnamefont {E.~J.}\ \bibnamefont {Mele}},\
  and\ \bibinfo {author} {\bibfnamefont {A.}~\bibnamefont {Vishwanath}},\
  }\href {https://doi.org/10.1103/RevModPhys.90.015001} {\bibfield  {journal}
  {\bibinfo  {journal} {Rev. Mod. Phys.}\ }\textbf {\bibinfo {volume} {90}},\
  \bibinfo {pages} {015001} (\bibinfo {year} {2018})}\BibitemShut {NoStop}%
\bibitem [{\citenamefont {Bzdu{\v{s}}ek}\ \emph {et~al.}(2016)\citenamefont
  {Bzdu{\v{s}}ek}, \citenamefont {Wu}, \citenamefont {R{\"u}egg}, \citenamefont
  {Sigrist},\ and\ \citenamefont {Soluyanov}}]{bzduvsek2016nodal}%
  \BibitemOpen
  \bibfield  {author} {\bibinfo {author} {\bibfnamefont {T.}~\bibnamefont
  {Bzdu{\v{s}}ek}}, \bibinfo {author} {\bibfnamefont {Q.}~\bibnamefont {Wu}},
  \bibinfo {author} {\bibfnamefont {A.}~\bibnamefont {R{\"u}egg}}, \bibinfo
  {author} {\bibfnamefont {M.}~\bibnamefont {Sigrist}},\ and\ \bibinfo {author}
  {\bibfnamefont {A.~A.}\ \bibnamefont {Soluyanov}},\ }\href@noop {} {\bibfield
   {journal} {\bibinfo  {journal} {Nature}\ }\textbf {\bibinfo {volume}
  {538}},\ \bibinfo {pages} {75} (\bibinfo {year} {2016})}\BibitemShut
  {NoStop}%
\bibitem [{\citenamefont {Bi}\ \emph {et~al.}(2017)\citenamefont {Bi},
  \citenamefont {Yan}, \citenamefont {Lu},\ and\ \citenamefont
  {Wang}}]{PhysRevB.96.201305}%
  \BibitemOpen
  \bibfield  {author} {\bibinfo {author} {\bibfnamefont {R.}~\bibnamefont
  {Bi}}, \bibinfo {author} {\bibfnamefont {Z.}~\bibnamefont {Yan}}, \bibinfo
  {author} {\bibfnamefont {L.}~\bibnamefont {Lu}},\ and\ \bibinfo {author}
  {\bibfnamefont {Z.}~\bibnamefont {Wang}},\ }\href
  {https://doi.org/10.1103/PhysRevB.96.201305} {\bibfield  {journal} {\bibinfo
  {journal} {Phys. Rev. B}\ }\textbf {\bibinfo {volume} {96}},\ \bibinfo
  {pages} {201305} (\bibinfo {year} {2017})}\BibitemShut {NoStop}%
\bibitem [{\citenamefont {Yan}\ \emph {et~al.}(2017)\citenamefont {Yan},
  \citenamefont {Bi}, \citenamefont {Shen}, \citenamefont {Lu}, \citenamefont
  {Zhang},\ and\ \citenamefont {Wang}}]{PhysRevB.96.041103}%
  \BibitemOpen
  \bibfield  {author} {\bibinfo {author} {\bibfnamefont {Z.}~\bibnamefont
  {Yan}}, \bibinfo {author} {\bibfnamefont {R.}~\bibnamefont {Bi}}, \bibinfo
  {author} {\bibfnamefont {H.}~\bibnamefont {Shen}}, \bibinfo {author}
  {\bibfnamefont {L.}~\bibnamefont {Lu}}, \bibinfo {author} {\bibfnamefont
  {S.-C.}\ \bibnamefont {Zhang}},\ and\ \bibinfo {author} {\bibfnamefont
  {Z.}~\bibnamefont {Wang}},\ }\href
  {https://doi.org/10.1103/PhysRevB.96.041103} {\bibfield  {journal} {\bibinfo
  {journal} {Phys. Rev. B}\ }\textbf {\bibinfo {volume} {96}},\ \bibinfo
  {pages} {041103} (\bibinfo {year} {2017})}\BibitemShut {NoStop}%
\bibitem [{\citenamefont {Yu}\ \emph {et~al.}(2017)\citenamefont {Yu},
  \citenamefont {Wu}, \citenamefont {Fang},\ and\ \citenamefont
  {Weng}}]{PhysRevLett.119.036401}%
  \BibitemOpen
  \bibfield  {author} {\bibinfo {author} {\bibfnamefont {R.}~\bibnamefont
  {Yu}}, \bibinfo {author} {\bibfnamefont {Q.}~\bibnamefont {Wu}}, \bibinfo
  {author} {\bibfnamefont {Z.}~\bibnamefont {Fang}},\ and\ \bibinfo {author}
  {\bibfnamefont {H.}~\bibnamefont {Weng}},\ }\href
  {https://doi.org/10.1103/PhysRevLett.119.036401} {\bibfield  {journal}
  {\bibinfo  {journal} {Phys. Rev. Lett.}\ }\textbf {\bibinfo {volume} {119}},\
  \bibinfo {pages} {036401} (\bibinfo {year} {2017})}\BibitemShut {NoStop}%
\bibitem [{\citenamefont {Yu}\ \emph {et~al.}(2015)\citenamefont {Yu},
  \citenamefont {Weng}, \citenamefont {Fang}, \citenamefont {Dai},\ and\
  \citenamefont {Hu}}]{PhysRevLett.115.036807}%
  \BibitemOpen
  \bibfield  {author} {\bibinfo {author} {\bibfnamefont {R.}~\bibnamefont
  {Yu}}, \bibinfo {author} {\bibfnamefont {H.}~\bibnamefont {Weng}}, \bibinfo
  {author} {\bibfnamefont {Z.}~\bibnamefont {Fang}}, \bibinfo {author}
  {\bibfnamefont {X.}~\bibnamefont {Dai}},\ and\ \bibinfo {author}
  {\bibfnamefont {X.}~\bibnamefont {Hu}},\ }\href
  {https://doi.org/10.1103/PhysRevLett.115.036807} {\bibfield  {journal}
  {\bibinfo  {journal} {Phys. Rev. Lett.}\ }\textbf {\bibinfo {volume} {115}},\
  \bibinfo {pages} {036807} (\bibinfo {year} {2015})}\BibitemShut {NoStop}%
\bibitem [{\citenamefont {Bzdušek}\ and\ \citenamefont
  {Sigrist}(2017)}]{PhysRevB.96.155105}%
  \BibitemOpen
  \bibfield  {author} {\bibinfo {author} {\bibfnamefont {T.}~\bibnamefont
  {Bzdušek}}\ and\ \bibinfo {author} {\bibfnamefont {M.}~\bibnamefont
  {Sigrist}},\ }\href {https://doi.org/10.1103/PhysRevB.96.155105} {\bibfield
  {journal} {\bibinfo  {journal} {Phys. Rev. B}\ }\textbf {\bibinfo {volume}
  {96}},\ \bibinfo {pages} {155105} (\bibinfo {year} {2017})}\BibitemShut
  {NoStop}%
\bibitem [{\citenamefont {Sun}\ \emph {et~al.}(2018)\citenamefont {Sun},
  \citenamefont {Zhang},\ and\ \citenamefont
  {Bzdušek}}]{PhysRevLett.121.106402}%
  \BibitemOpen
  \bibfield  {author} {\bibinfo {author} {\bibfnamefont {X.-Q.}\ \bibnamefont
  {Sun}}, \bibinfo {author} {\bibfnamefont {S.-C.}\ \bibnamefont {Zhang}},\
  and\ \bibinfo {author} {\bibfnamefont {T.}~\bibnamefont {Bzdušek}},\ }\href
  {https://doi.org/10.1103/PhysRevLett.121.106402} {\bibfield  {journal}
  {\bibinfo  {journal} {Phys. Rev. Lett.}\ }\textbf {\bibinfo {volume} {121}},\
  \bibinfo {pages} {106402} (\bibinfo {year} {2018})}\BibitemShut {NoStop}%
\bibitem [{\citenamefont {Tiwari}\ and\ \citenamefont
  {Bzdušek}(2020)}]{PhysRevB.101.195130}%
  \BibitemOpen
  \bibfield  {author} {\bibinfo {author} {\bibfnamefont {A.}~\bibnamefont
  {Tiwari}}\ and\ \bibinfo {author} {\bibfnamefont {T.}~\bibnamefont
  {Bzdušek}},\ }\href {https://doi.org/10.1103/PhysRevB.101.195130} {\bibfield
   {journal} {\bibinfo  {journal} {Phys. Rev. B}\ }\textbf {\bibinfo {volume}
  {101}},\ \bibinfo {pages} {195130} (\bibinfo {year} {2020})}\BibitemShut
  {NoStop}%
\bibitem [{\citenamefont {Bouhon}\ \emph {et~al.}(2020)\citenamefont {Bouhon},
  \citenamefont {Wu}, \citenamefont {Slager}, \citenamefont {Weng},
  \citenamefont {Yazyev},\ and\ \citenamefont {Bzdu{\v{s}}ek}}]{bouhon2020non}%
  \BibitemOpen
  \bibfield  {author} {\bibinfo {author} {\bibfnamefont {A.}~\bibnamefont
  {Bouhon}}, \bibinfo {author} {\bibfnamefont {Q.}~\bibnamefont {Wu}}, \bibinfo
  {author} {\bibfnamefont {R.-J.}\ \bibnamefont {Slager}}, \bibinfo {author}
  {\bibfnamefont {H.}~\bibnamefont {Weng}}, \bibinfo {author} {\bibfnamefont
  {O.~V.}\ \bibnamefont {Yazyev}},\ and\ \bibinfo {author} {\bibfnamefont
  {T.}~\bibnamefont {Bzdu{\v{s}}ek}},\ }\href@noop {} {\bibfield  {journal}
  {\bibinfo  {journal} {Nat. Phys.}\ }\textbf {\bibinfo {volume} {16}},\
  \bibinfo {pages} {1137} (\bibinfo {year} {2020})}\BibitemShut {NoStop}%
\bibitem [{\citenamefont {Jiang}\ \emph {et~al.}(2021)\citenamefont {Jiang},
  \citenamefont {Bouhon}, \citenamefont {Lin}, \citenamefont {Zhou},
  \citenamefont {Hou}, \citenamefont {Li}, \citenamefont {Slager},\ and\
  \citenamefont {Jiang}}]{jiang2021experimental}%
  \BibitemOpen
  \bibfield  {author} {\bibinfo {author} {\bibfnamefont {B.}~\bibnamefont
  {Jiang}}, \bibinfo {author} {\bibfnamefont {A.}~\bibnamefont {Bouhon}},
  \bibinfo {author} {\bibfnamefont {Z.-K.}\ \bibnamefont {Lin}}, \bibinfo
  {author} {\bibfnamefont {X.}~\bibnamefont {Zhou}}, \bibinfo {author}
  {\bibfnamefont {B.}~\bibnamefont {Hou}}, \bibinfo {author} {\bibfnamefont
  {F.}~\bibnamefont {Li}}, \bibinfo {author} {\bibfnamefont {R.-J.}\
  \bibnamefont {Slager}},\ and\ \bibinfo {author} {\bibfnamefont {J.-H.}\
  \bibnamefont {Jiang}},\ }\href@noop {} {\bibfield  {journal} {\bibinfo
  {journal} {Nat. Phys}\ }\textbf {\bibinfo {volume} {17}},\ \bibinfo {pages}
  {1239} (\bibinfo {year} {2021})}\BibitemShut {NoStop}%
\bibitem [{\citenamefont {Guo}\ \emph {et~al.}(2021)\citenamefont {Guo},
  \citenamefont {Jiang}, \citenamefont {Zhang}, \citenamefont {Zhang},
  \citenamefont {Zhang}, \citenamefont {Yang}, \citenamefont {Zhang},\ and\
  \citenamefont {Chan}}]{guo2021experimental}%
  \BibitemOpen
  \bibfield  {author} {\bibinfo {author} {\bibfnamefont {Q.}~\bibnamefont
  {Guo}}, \bibinfo {author} {\bibfnamefont {T.}~\bibnamefont {Jiang}}, \bibinfo
  {author} {\bibfnamefont {R.-Y.}\ \bibnamefont {Zhang}}, \bibinfo {author}
  {\bibfnamefont {L.}~\bibnamefont {Zhang}}, \bibinfo {author} {\bibfnamefont
  {Z.-Q.}\ \bibnamefont {Zhang}}, \bibinfo {author} {\bibfnamefont
  {B.}~\bibnamefont {Yang}}, \bibinfo {author} {\bibfnamefont {S.}~\bibnamefont
  {Zhang}},\ and\ \bibinfo {author} {\bibfnamefont {C.~T.}\ \bibnamefont
  {Chan}},\ }\href@noop {} {\bibfield  {journal} {\bibinfo  {journal} {Nature}\
  }\textbf {\bibinfo {volume} {594}},\ \bibinfo {pages} {195} (\bibinfo {year}
  {2021})}\BibitemShut {NoStop}%
\bibitem [{\citenamefont {Yan}\ \emph {et~al.}(2018)\citenamefont {Yan},
  \citenamefont {Liu}, \citenamefont {Yan}, \citenamefont {Liu}, \citenamefont
  {Chen}, \citenamefont {Wang},\ and\ \citenamefont
  {Lu}}]{yan2018experimental}%
  \BibitemOpen
  \bibfield  {author} {\bibinfo {author} {\bibfnamefont {Q.}~\bibnamefont
  {Yan}}, \bibinfo {author} {\bibfnamefont {R.}~\bibnamefont {Liu}}, \bibinfo
  {author} {\bibfnamefont {Z.}~\bibnamefont {Yan}}, \bibinfo {author}
  {\bibfnamefont {B.}~\bibnamefont {Liu}}, \bibinfo {author} {\bibfnamefont
  {H.}~\bibnamefont {Chen}}, \bibinfo {author} {\bibfnamefont {Z.}~\bibnamefont
  {Wang}},\ and\ \bibinfo {author} {\bibfnamefont {L.}~\bibnamefont {Lu}},\
  }\href@noop {} {\bibfield  {journal} {\bibinfo  {journal} {Nature Physics}\
  }\textbf {\bibinfo {volume} {14}},\ \bibinfo {pages} {461} (\bibinfo {year}
  {2018})}\BibitemShut {NoStop}%
\bibitem [{\citenamefont {Yang}\ \emph {et~al.}(2020)\citenamefont {Yang},
  \citenamefont {Yang}, \citenamefont {You}, \citenamefont {Chan},
  \citenamefont {Mao}, \citenamefont {Guo}, \citenamefont {Ma}, \citenamefont
  {Xia}, \citenamefont {Fan}, \citenamefont {Xiang},\ and\ \citenamefont
  {Zhang}}]{PhysRevLett.125.033901}%
  \BibitemOpen
  \bibfield  {author} {\bibinfo {author} {\bibfnamefont {E.}~\bibnamefont
  {Yang}}, \bibinfo {author} {\bibfnamefont {B.}~\bibnamefont {Yang}}, \bibinfo
  {author} {\bibfnamefont {O.}~\bibnamefont {You}}, \bibinfo {author}
  {\bibfnamefont {H.-C.}\ \bibnamefont {Chan}}, \bibinfo {author}
  {\bibfnamefont {P.}~\bibnamefont {Mao}}, \bibinfo {author} {\bibfnamefont
  {Q.}~\bibnamefont {Guo}}, \bibinfo {author} {\bibfnamefont {S.}~\bibnamefont
  {Ma}}, \bibinfo {author} {\bibfnamefont {L.}~\bibnamefont {Xia}}, \bibinfo
  {author} {\bibfnamefont {D.}~\bibnamefont {Fan}}, \bibinfo {author}
  {\bibfnamefont {Y.}~\bibnamefont {Xiang}},\ and\ \bibinfo {author}
  {\bibfnamefont {S.}~\bibnamefont {Zhang}},\ }\href
  {https://doi.org/10.1103/PhysRevLett.125.033901} {\bibfield  {journal}
  {\bibinfo  {journal} {Phys. Rev. Lett.}\ }\textbf {\bibinfo {volume} {125}},\
  \bibinfo {pages} {033901} (\bibinfo {year} {2020})}\BibitemShut {NoStop}%
\bibitem [{\citenamefont {Ningyuan}\ \emph {et~al.}(2015)\citenamefont
  {Ningyuan}, \citenamefont {Owens}, \citenamefont {Sommer}, \citenamefont
  {Schuster},\ and\ \citenamefont {Simon}}]{PhysRevX.5.021031}%
  \BibitemOpen
  \bibfield  {author} {\bibinfo {author} {\bibfnamefont {J.}~\bibnamefont
  {Ningyuan}}, \bibinfo {author} {\bibfnamefont {C.}~\bibnamefont {Owens}},
  \bibinfo {author} {\bibfnamefont {A.}~\bibnamefont {Sommer}}, \bibinfo
  {author} {\bibfnamefont {D.}~\bibnamefont {Schuster}},\ and\ \bibinfo
  {author} {\bibfnamefont {J.}~\bibnamefont {Simon}},\ }\href
  {https://doi.org/10.1103/PhysRevX.5.021031} {\bibfield  {journal} {\bibinfo
  {journal} {Phys. Rev. X}\ }\textbf {\bibinfo {volume} {5}},\ \bibinfo {pages}
  {021031} (\bibinfo {year} {2015})}\BibitemShut {NoStop}%
\bibitem [{\citenamefont {Lee}\ \emph {et~al.}(2020)\citenamefont {Lee},
  \citenamefont {Sutrisno}, \citenamefont {Hofmann}, \citenamefont {Helbig},
  \citenamefont {Liu}, \citenamefont {Ang}, \citenamefont {Ang}, \citenamefont
  {Zhang}, \citenamefont {Greiter},\ and\ \citenamefont
  {Thomale}}]{lee2020imaging}%
  \BibitemOpen
  \bibfield  {author} {\bibinfo {author} {\bibfnamefont {C.~H.}\ \bibnamefont
  {Lee}}, \bibinfo {author} {\bibfnamefont {A.}~\bibnamefont {Sutrisno}},
  \bibinfo {author} {\bibfnamefont {T.}~\bibnamefont {Hofmann}}, \bibinfo
  {author} {\bibfnamefont {T.}~\bibnamefont {Helbig}}, \bibinfo {author}
  {\bibfnamefont {Y.}~\bibnamefont {Liu}}, \bibinfo {author} {\bibfnamefont
  {Y.~S.}\ \bibnamefont {Ang}}, \bibinfo {author} {\bibfnamefont {L.~K.}\
  \bibnamefont {Ang}}, \bibinfo {author} {\bibfnamefont {X.}~\bibnamefont
  {Zhang}}, \bibinfo {author} {\bibfnamefont {M.}~\bibnamefont {Greiter}},\
  and\ \bibinfo {author} {\bibfnamefont {R.}~\bibnamefont {Thomale}},\
  }\href@noop {} {\bibfield  {journal} {\bibinfo  {journal} {Nature
  communications}\ }\textbf {\bibinfo {volume} {11}},\ \bibinfo {pages} {1}
  (\bibinfo {year} {2020})}\BibitemShut {NoStop}%
\bibitem [{wu()}]{wu}%
  \BibitemOpen
  \href@noop {} {\bibinfo  {journal} {See Supplemental Material}\ }\BibitemShut
  {NoStop}%
\bibitem [{\citenamefont {Yang}\ \emph {et~al.}(2015)\citenamefont {Yang},
  \citenamefont {Gao}, \citenamefont {Shi}, \citenamefont {Lin}, \citenamefont
  {Gao}, \citenamefont {Chong},\ and\ \citenamefont
  {Zhang}}]{PhysRevLett.114.114301}%
  \BibitemOpen
\bibfield  {journal} {  }\bibfield  {author} {\bibinfo {author} {\bibfnamefont
  {Z.}~\bibnamefont {Yang}}, \bibinfo {author} {\bibfnamefont {F.}~\bibnamefont
  {Gao}}, \bibinfo {author} {\bibfnamefont {X.}~\bibnamefont {Shi}}, \bibinfo
  {author} {\bibfnamefont {X.}~\bibnamefont {Lin}}, \bibinfo {author}
  {\bibfnamefont {Z.}~\bibnamefont {Gao}}, \bibinfo {author} {\bibfnamefont
  {Y.}~\bibnamefont {Chong}},\ and\ \bibinfo {author} {\bibfnamefont
  {B.}~\bibnamefont {Zhang}},\ }\href
  {https://doi.org/10.1103/PhysRevLett.114.114301} {\bibfield  {journal}
  {\bibinfo  {journal} {Phys. Rev. Lett.}\ }\textbf {\bibinfo {volume} {114}},\
  \bibinfo {pages} {114301} (\bibinfo {year} {2015})}\BibitemShut {NoStop}%
\bibitem [{\citenamefont {Zhao}\ \emph {et~al.}(2009)\citenamefont {Zhao},
  \citenamefont {Zhou}, \citenamefont {Zhang},\ and\ \citenamefont
  {Lippens}}]{ZHAO200960}%
  \BibitemOpen
  \bibfield  {author} {\bibinfo {author} {\bibfnamefont {Q.}~\bibnamefont
  {Zhao}}, \bibinfo {author} {\bibfnamefont {J.}~\bibnamefont {Zhou}}, \bibinfo
  {author} {\bibfnamefont {F.}~\bibnamefont {Zhang}},\ and\ \bibinfo {author}
  {\bibfnamefont {D.}~\bibnamefont {Lippens}},\ }\href
  {https://doi.org/https://doi.org/10.1016/S1369-7021(09)70318-9} {\bibfield
  {journal} {\bibinfo  {journal} {Materials Today}\ }\textbf {\bibinfo {volume}
  {12}},\ \bibinfo {pages} {60} (\bibinfo {year} {2009})}\BibitemShut {NoStop}%
\end{thebibliography}%
\end{document}